\documentclass[a4paper,10pt]{scrartcl}

\usepackage{latexsym}
\usepackage{amsmath}
\usepackage{amsfonts}
\usepackage{amssymb}
\usepackage{graphicx}
\usepackage[english]{babel}
\usepackage{wasysym}
\usepackage{units}
\newcommand{\p}{\partial}
\newcommand{\reff}[1]{(\ref{#1})}
\newcommand{\vs}[1]{\vspace{#1mm}}

\newcommand{\vsO}{\vspace{.1cm}\hfill\\}
\newcommand{\vsT}{\vspace{.2cm}\hfill\\}

\title{\Large THE JEANS INSTABILITY\\ IN PRESENCE OF VISCOUS EFFECTS}

\author{Nakia Carlevaro$^{\;a,\;b}$ and Giovanni Montani$^{\;b,\;c,\;d,\;e}$\vsT
\emph{\footnotesize $^a$Department of Physics, Polo Scientifico -- Universit\`a degli Studi di Firenze,}\vs{-2.5}\\
\emph{\footnotesize INFN -- Section of Florence, Via G. Sansone, 1 (50019), Sesto Fiorentino (FI), Italy}\\
\emph{\footnotesize $^b$ICRA -- International Center for Relativistic Astrophysics,}\vs{-2.5}\\
\emph{\footnotesize c/o Dep. of Physics - ``Sapienza'' Universit\`a di Roma}\\
\emph{\footnotesize $^c$ Department of Physics - ``Sapienza'' Universit\`a di Roma, Piazza A. Moro, 5 (00185), Rome, Italy}\\
\emph{\footnotesize $^d$ENEA -- C.R. Frascati (Department F.P.N.), Via Enrico Fermi, 45 (00044), Frascati (Rome), Italy}\\
\emph{\footnotesize $^{e}$ ICRANet -- C. C. Pescara, Piazzale della Repubblica, 10 (65100), Pescara, Italy}\vsO
{\footnotesize\ttfamily nakia.carlevaro@icra.it\quad montani@icra.it}
}
\date{}
\begin{document}
\maketitle

%
\hrule
\begin{abstract} \textbf{Abstract:} An analysis of the gravitational instability in presence of dissipative effects is addressed. In particular, the standard Jeans Mechanism and the generalization in treating the Universe expansion are both analyzed when bulk viscosity affects the first-order Newtonian dynamics. As results, the perturbation evolution is founded to be damped by dissipative processes and the top-down mechanism of structure fragmentation is suppressed. In such a scheme, the value of the Jeans Mass remains unchanged also in presence of viscosity.
\end{abstract}
\hrule

\vspace{1cm}
\section{Introduction}
The \emph{Cosmological Standard Model} \cite{kolb-turner} well describes many parts of the Universe evolution and it takes into account the Friedmann-Lema\^{i}tre-Robertson-Walker (FLRW) metric as the highest symmetric background. In fact, considering the mean energy-density at big scales, \emph{i.e.}, greater than $100$ $Mpc$, it tends to an homogeneous distribution. On the other hand, the Universe observation at small scales shows a very inhomogeneous and anisotropic matter- and energy-distribution. In this respect, the isotropic hypothesis of the \emph{Cosmological Principle} \cite{weinberg} is not based on the big-scale observations but on the strong isotropy of the Cosmic Background Radiation, which has a black-body spectral-distribution at $T\sim2.73\,$K with temperature fluctuations of order $\mathcal{O}(10^{-4})$. Moreover, the Cosmological Standard Model is confirmed by the primordial-nucleosynthesis prediction for the light elements, which is in agreement with direct observations. 

Such a dichotomy between the isotropy of region at red-shift $z_{rs}\sim10^{3}$ and the extreme irregularity of the recent Universe, $z_{rs}\ll1$, is at the ground of the interest in the study of the gravitational instability for the structure formation. The study of the cosmological perturbation evolution can be separated in two distinct regimes: the linear regime, concerning \emph{density contrast} $\delta$ much less than the unity, and the non-linear one in which $\delta>1$, giving rise to the effective structure formation. Despite the approximate hypotheses, the linear regime provides interesting predictive information also at low red-shift, since an analytical description can be addressed to study the growth of the density contrast. 

As matter of fact, we underline that the study of the perturbation dynamics in the radiation-dominated Early Universe requires a pure relativistic treatment, in order to correlate the matter fluctuations with the geometrical ones \cite{quasi-iso,flrw-nak}. On the other hand, the evolution during the matter-dominated era can be consistently described using the Newtonian approximation picture, as soon as sub-horizon-sized scales are treated. In this scheme, the fundamental result of the density-perturbation analysis is the so-called \emph{Jeans Mass}, which is the threshold value for the fluctuation masses to condense generating a real structure. If masses greater than the Jeans Mass are addressed, density perturbations begin to diverge as function of time giving rise to the gravitational collapse \cite{jeans02,jeans}. 

In this work, we are aimed to consider dissipative effects into the fluid dynamics, in the linear Newtonian regime. The starting point is the Eulerian set of equations governing the fluid motion (Section 2) on which one can develop a perturbative theory by adding small fluctuations to the unperturbed background (see also \cite{corona,garcia}). In particular, we introduce in the first-order analysis the so-called \emph{bulk viscosity} (we neglect the shear or first viscosity since we are dealing with homogeneous model and no internal frictions arise). Such kind of viscosity can be expressed in terms of the thermodynamical parameters of the fluid. Following the line of the ``Landau School'' \cite{bk76,bk77,bnk79,lk63,l46}, we phenomenologically describe this quantity as a function of the Universe energy density $\rho$ via a power-law: $\zeta=z\,\rho^{\,s}$ where $s=const.$
and $z$ is a parameter which defines the intensity of viscous effects, see also  \cite{b87,b88,paddy}. 

Two different cases are treated: the standard \emph{Jeans Mechanism} \cite{jeans02,weinberg} and the generalization in presence of the expanding Universe \cite{weinberg,w71}. In the first approach (Section 3), the unperturbed background is assumed to be characterized by a static and uniform solution of the fluid parameters while, in the second case (Section 4), we consider the effects of the Universe expansion on the zeroth-order dynamics which is now described by the \emph{Friedmann Eq.'s} for an homogeneous and isotropic Universe. 

As a result, we show how the presence of bulk viscosity damps the density-contrast growth, suppressing the structure formation in the \emph{top-down fragmentation scheme} \cite{kolb-turner} (mainly associated with the hot dark-matter phenomenology  \cite{capo2,max,odintsov1,odintsov2}) without changing the threshold value of the Jeans Mass. In fact, in both cases, the grater the viscosity coefficient is, the slower the density contrast diverges in time, in the limit $t\to\infty$. In this respect, considering a collapsing macro-structure, \emph{i.e.}, of mass greater than the Jeans Mass, if viscosity becomes sufficiently large then the Jeans fragmentation mechanism is deeply unfavored since the sub-scale density-contrasts are damped. In particular, in treating the Jeans Model, bulk viscosity affects the perturbation evolution generating a new decreasing regime in place of the pure oscillatory behavior addressed in the Jeans analysis.

The main merit of this work is to be determined in having traced a possible scenario for fragmentation processes in presence of viscosity. We infer that the unfavored nature of the top-down mechanism, appearing when a viscous trace is present, can survive also in the non-linear regime when dissipative effects play surely an important role in the structure formation.

\newcommand{\vo}{\textbf{v}_{\scriptscriptstyle 0}}
\newcommand{\ro}{\rho_{\scriptscriptstyle 0}}
\newcommand{\rb}{\bar{\rho}}
\newcommand{\gb}{\bar{\gamma}}
\newcommand{\wb}{\bar{\omega}}
\newcommand{\po}{p_{\scriptscriptstyle 0}}
\newcommand{\phio}{\phi_{\scriptscriptstyle 0}}
\newcommand{\zetao}{\zeta_{\scriptscriptstyle 0}}
\newcommand{\zetaob}{\bar{\zeta}_{\scriptscriptstyle 0}}
\newcommand{\w}{\omega}
\newcommand{\ao}{a_{\scriptscriptstyle 0}}
\newcommand{\vu}{\textbf{v}_{\scriptscriptstyle 1}}
\newcommand{\ru}{\rho_{\scriptscriptstyle 1}}
\newcommand{\pu}{p_{\scriptscriptstyle 1}}
\newcommand{\phiu}{\phi_{\scriptscriptstyle 1}}
\newcommand{\zetau}{\zeta_{\scriptscriptstyle 1}}
\newcommand{\vup}{\dot{\textbf{v}}_{\scriptscriptstyle 1}}
\newcommand{\rup}{\dot{\rho}_{\scriptscriptstyle 1}}

\section{Motion Equations of Viscous Fluids}
In order to describe the Newtonian evolution of a fluid, we here want to introduce the Eulerian equations governing the fluid parameters: the density $\rho$, the local 3-velocity $\textbf{v}$ (of components $v_\alpha$) and the pressure $p$, in presence of a gravitational potential $\phi$.

Adiabatic \emph{ideal} fluids are governed, in Newtonian regime, by the following set of equations \cite{landau-fluid}: the \emph{Continuity Eq.}, which guarantees the energy conservation
\begin{equation}
\p_t{\rho}+\nabla\cdot(\rho\textbf{v})=0\;,
\end{equation}
the \emph{Euler Eq.}, which ensures the momentum conservation
\begin{equation}
\rho\,\p_t{\textbf{v}}+\rho\,(\textbf{v}\cdot{\nabla})\textbf{v}=
-{\nabla}p-\rho\,{\nabla}\phi\;,
\end{equation}
while pressure and density are linked by the \emph{Eq. of State} (EoS): $p=p\,(\rho)$. In this picture, the sound speed is defined by the relation $v_s^2=\delta p/\delta\rho$.

Let us introduce the effects of the energy dissipation during the motion of the fluid, due to the thermodynamical non-reversibility and to internal friction (we neglect the thermal conductivity). To obtain the motion equations for a viscous fluid, we have to include some  additional terms in the ideal fluid description. The Continuity Eq. is derived by the time evolution of the matter density and by the mass conservation law. This way, it remains valid for any kind of fluid. Euler Eq., in absence of the gravitational field, rewrites (here $\alpha=1,2,3$)
\begin{equation}
\p_t(\rho v_\alpha)=-\p_\beta\,\Pi_{\alpha\beta}\;,
\end{equation}
where $\Pi_{\alpha\beta}$ denotes the momentum-flux energy-tensor. If ideal fluids are addressed, we deal with completely reversible transfer of momentum, obtaining the expression: $\Pi_{\alpha\beta}=p\,\delta_{\alpha\beta}\,+\,\rho\,v_\alpha v_\beta$. Viscosity is responsible for an additional term $\tilde{\sigma}_{\alpha\beta}$ to this expression due to another irreversible momentum transfer, where non-vanishing velocity gradients are present. For a viscous fluid we get  
\begin{equation}
\Pi_{\alpha\beta}=p\,\delta_{\alpha\beta}\,+\,\rho\,v_\alpha v_\beta\,
-\,\tilde{\sigma}_{\alpha\beta}\,
=\,-\sigma_{\alpha\beta}\,+\,\rho\,v_\alpha v_\beta\;,
\qquad
\sigma_{\alpha\beta}=-p\,\delta_{\alpha\beta}\,+\,\tilde{\sigma}_{\alpha\beta}\;,
\end{equation}
where $\sigma_{\alpha\beta}$ is the \emph{stress tensor} and $\tilde{\sigma}_{\alpha\beta}$ is called the \emph{viscous stress-tensor}.

The general form of $\tilde{\sigma}_{\alpha\beta}$ can be derived by a qualitative analysis of the velocity gradients in presence of uniform rotation and volume changes of the fluid. The most general form of the viscous stress tensor is \cite{landau-fluid} 
\begin{equation}
\tilde{\sigma}_{\alpha\beta}=\,\eta\;(\p_\beta v_\alpha+\p_\alpha v_\beta
-\tfrac{2}{3}\,\delta_{\alpha\beta}\p_\gamma v_\gamma)
+\zeta\,\delta_{\alpha\beta}\,\p_\gamma v_\gamma\;,
\end{equation}
where the coefficients $\eta$ e $\zeta$ are not dependent of velocity (the fluid is isotropic and its properties must be described only by scalar quantities) and the term proportional to $\eta$ coefficient vanishes for the $\alpha$ and $\beta$ contraction. Here, the coefficient $\eta$ is called \emph{shear viscosity} while $\zeta$ denotes bulk viscosity and they are both positive quantities.

Using Continuity Eq., the ideal fluid Euler Eq. rewrites
\begin{equation}\nonumber
\rho(\p_t{v}_\alpha + v_\beta\,\p_\beta v_\alpha)\,
=\,-\p_\alpha p\;,
\end{equation}
and the motion equation of a viscous fluids can now be obtained by adding the expression $\p_\beta\tilde{\sigma}_{\alpha\beta}$ to the rhs of the equation above, obtaining
\begin{align}\nonumber
\rho(\p_t{v}_\alpha + v_\beta\,\p_\beta v_\alpha)=
-\p_\alpha p\,+\,\p_\beta[\eta\;(\p_\beta v_\alpha+\p_\alpha v_\beta
-\tfrac{2}{3}\,\delta_{\alpha\beta}\p_\gamma v_\gamma)]\,+\,
\p_\alpha(\zeta\,\p_\gamma v_\gamma)\;.
\end{align}

The viscous coefficients are not constant and we have to express their dependence on the state parameters of the fluid. Since we are interested to treat isotropic and homogeneous perturbative cosmological models, we can safely neglect the first viscosity (shear viscosity) in the unperturbed dynamics. In fact, in such models there is no displacement of matter layers with respect to each other and this kind of viscosity represents the energy dissipation due to this effect. Indeed, in presence of small inhomogeneities such affect should be taken into account, in principle. However, in this work, we are aimed at studying the behavior of scalar density perturbations. In this respect, volume changes of a given mass scale are essentially involved and, therefore, we concentrate our attention to bulk-viscosity effects only. In fact, we expect that the non-equilibrium dynamics of matter compression and rarefaction is more relevant than friction among the different layers.

According to literature developments \cite{bk76,bk77,bnk79,b87,paddy}, we now assume the bulk-viscosity coefficient as a function of the energy density $\rho(t)$ expressed via a power-law of the form
\begin{equation}\label{bulk-power-law}
\zeta=z\,\rho^{\,s}\;,
\end{equation}
where $s=const.$ and $z$ is a constant parameter which defines the intensity of viscous effects. With these assumptions, the Euler Eq. takes the following form 
\begin{equation}
\rho\,\p_t{\textbf{v}}+ \rho\,(\textbf{v}\cdot\nabla)\textbf{v} + \nabla p\,-
\zeta\,\nabla(\nabla \cdot \textbf{v})=0\;,
\end{equation}
which is the well-known \emph{Navier-Stokes Eq}.

This analysis is developed without considering the gravitational field, which has to be introduced in the Euler Eq. as usual. We have also to consider the equation describing the gravitational field itself: the \emph{Poisson Eq}. Let us now recall the set of motion equations in the case of an adiabatic viscous fluid:
\begin{subequations}\label{newtonian-motion-eq}
\begin{align}
\label{continuity-eq}
\p_t{\rho}+\nabla\cdot(\rho\textbf{v})&=0\;,\\
\label{navier-stokes-eq}
\rho\,\p_t{\textbf{v}}+ \rho\,(\textbf{v}\cdot\nabla)\textbf{v} + \nabla p\,-
\zeta\,\nabla(\nabla \cdot \textbf{v})+\rho\,\nabla\phi&=0\;,\\
\label{poisson-eq}
\nabla^2 \phi-4\pi G \rho &=0\;,
\end{align}
\end{subequations}
such a system is the starting point to analyze the gravitational instability.

\section{Analysis of the dissipative Jeans Mechanism}
The Universe is uniform at big scales but many concentrations are presented at small scales, \emph{e.g.}, galaxies and clusters, where the mass density is larger than the Universe mean-density. These mass agglomerates are due to the gravitational instability: if density perturbations are generated in a certain volume, the gravitational forces act contracting this volume, allowing a gravitational collapse. The only forces which contrast such gravitational contraction are the pressure ones, which act in order to maintain uniform the energy density. The Jeans Mechanism analyzes what are the conditions for which density perturbations become unstable to the gravitational collapse. 

This model \cite{jeans02} is based on a Newtonian approach and the effects of the expanding Universe are neglected. The fundamental hypothesis of such an analysis is a static and uniform solution for the zeroth-order dynamics
\begin{equation}\label{static-swindle-solution}
\vo=0\;,\qquad\ro=const.\;,\qquad \po=const.\;,\qquad\phio=const.
\end{equation}
Of course, this assumption contradicts the gravitational equation, but we follow the original Jeans analysis imposing the so-called ``Jeans swindle'' \cite{jeans,weinberg}. We underline that our study will focus on Universe stages when the mean density is very small: in particular the recombination era, after decoupling. This way, the effects of bulk viscosity on the unperturbed dynamics can be consistently neglected in view of its phenomenological behavior \reff{bulk-power-law}.

\subsection{Remarks on the Jeans Model}
In the standard Jeans Model, a perfect fluid background is assumed. After setting $\zeta=0$ in the motion equations \reff{newtonian-motion-eq}, let us add small fluctuations to the unperturbed solution: $\rho=\ro+\delta\rho,\;p=\po+\delta p,\;\phi=\phio+\delta\phi,\;\textbf{v}=\vo+\delta\textbf{v}$. Furthermore, only adiabatic perturbations are treated and the sound speed is defined as $v_s^2=\p p/\p\rho$. Substituting such expressions in the system \reff{newtonian-motion-eq}, after standard manipulation (we remark that second-order terms are neglected), one differential equation for the density perturbations can be derived:
\begin{equation}\label{eqdiffjeans}
\p_t^2{\delta\rho}-v_s^2\;\nabla^2\delta\rho=4\pi G\,\ro\,\delta\rho\;.
\end{equation}
To study the properties of $\delta\rho$, we now consider a plane-wave solutions of the form
\begin{equation}\label{plane-wave}	
\delta\rho\,(\textbf{r},t) = A\;e^{i\omega t - i\textbf{k}\cdot\textbf{r}}\;,
\end{equation}
where $\omega$ and $\textbf{k}$ ($k=|\textbf{k}|$) are the angular frequency and the wave number, respectively. This way, one can obtain the following dispersion relation
\begin{equation}\label{dispersion-jeans}
\omega^2=v_s^2k^2-4\pi G\,{\ro}\;.
\end{equation}

In this scheme, two different regimes are present: if $\omega^{2}>0$ a pure time oscillatory-behavior for density perturbations is obtained. While if $\omega^{2}<0$, the fluctuations exponentially grow in time, in the $t\to\infty$ asymptotic limit (\emph{i.e.}, we choose the negative imaginary part of the angular frequency solution) and the gravitational collapse is addressed since also the density contrast $\delta=\delta\rho/\ro$ diverges. The condition $\omega^2=0$ defines the so-called \emph{Jeans Scale} $K_J$ and the Jeans Mass $M_J$ (which is the total mass in a sphere of radius $R=\pi/K_J$). Such threshold quantities read
\begin{equation}\label{jeans-mass}
K_J=\rho_{\scriptscriptstyle 0}\sqrt{\frac{4\pi G
\rho_{\scriptscriptstyle 0}}{v_s^2}}\;,\qquad
M_J=\tfrac{4}{3}\pi\left(\frac{2\pi}{K_J}\right)^3
\rho_{\scriptscriptstyle 0}=\tfrac{4}{3}\,
\frac{\pi^{5/2}\,v_s^3}{G^{3/2}{\rho_{\scriptscriptstyle 0}}^{1/2}}\;.
\end{equation}

Let us now analyze in some details the two regimes. In the case $M<M_J$ (\emph{i.e.}, $\w^2>0$), $\delta\rho$ behave like two progressive sound waves (with constant amplitude) propagating in the $\pm\textbf{k}$ directions with velocity $v_w=v_s(1-(K_J/k)^2)^{1/2}$. In the limit $k\to\infty$, the propagation velocity approaches the value $v_s$, and fluctuations behave like pure sound waves. On the other hand, if $k\to K_J$, stationary waves are addressed (\emph{i.e.}, $v_w=0$). 

In the case $M>M_J$ (\emph{i.e.}, $\w^2<0$), density perturbations evolve like stationary waves with a time dependent amplitude. In particular, choosing the negative imaginary part of the solution for $\w$, the wave amplitude exponentially explodes, generating the gravitational collapse.

\subsection{Jeans Mechanism in presence of bulk viscosity}
Let us now analyze how viscosity can affect the gravitational-collapse dynamics. We recall that the only viscous process which can be addressed in an homogeneous and isotropic model is bulk viscosity. As discussed at the beginning of this Section, we are able to neglect such kind of viscosity in the unperturbed dynamics, which results to be described by the static and uniform solution \reff{static-swindle-solution}. 

We now start by adding the usual small fluctuations to such a solution, \emph{i.e.}, $\delta\rho,\;\delta p,\;\delta\phi,\;\delta\textbf{v}$. In treating bulk-viscosity perturbations, we use the expansion $\zeta=\zetao+\delta\zeta$ where 
\begin{equation}\label{bulk-expansion}
\zetao=\zeta(\ro)=z\ro^{s}=const.\;,\qquad
\delta\zeta=\delta\rho\;(\p\zeta/\p \rho)+...=
\,z\,s\,\ro^{\,s-1}\,\delta\rho\;+...\;.
\end{equation}
Substituting all fluctuations in the system \reff{newtonian-motion-eq}, we get the first-order motion-equations of the model
\begin{subequations}\label{system-first-order}
\begin{align}
\label{continuitapert3.1}
\p_t{\delta\rho}+\ro\nabla\cdot\delta\textbf{v}&=0\;,\\ 
\label{euleropert3.1}
\ro\,\p_t{\delta\textbf{v}}+v_s^2\,\nabla\delta\rho+\ro\,\nabla\delta\phi-
\zetao\,\nabla(\nabla \cdot \delta\textbf{v})&=0\;,\\
\label{campogpert3.1}
\nabla^2 \delta \phi - 4\pi G\,\delta\rho&=0\;.
\end{align}
\end{subequations}
With some little algebra, one can obtain an unique equation for density perturbations, describing the dynamics of the gravitational collapse:
\begin{equation}\label{eqfond3.1}
\ro\;\p_t^2{\delta\rho}-\ro\;v_s^2\nabla^2\,\delta\rho-
\zetao\,\nabla^2\,\p_t{\delta\rho}=4\pi G\ro^2\;\delta\rho\;.
\end{equation}

Using the linearity of the equation above, a decomposition in Fourier expansion can be performed. This way, plane waves solutions \reff{plane-wave} can be addressed, obtaining a generalized dispersion relation
\begin{equation}
\ro\,\omega^2-\,i\,\zetao\,k^2\;\omega+\ro(4\pi G\ro-v_s^2 k^2)=0\;.
\end{equation}

As in the standard Jean Model, the nature of the angular frequency is responsible of two different regimes for the density-perturbation evolution. The dispersion relation has the solution 
\begin{equation}
\omega=i\,\frac{{\zetao} k^2}{2\,{\ro}}\pm
\sqrt{\bar{\omega}}\;,\qquad\;\;
\bar{\omega}=-\frac{k^4{\zetao}^2}{4{\ro}^2}+v_s^2k^2-4\pi G{\ro}\;,
\end{equation}
thus we obtain the time exponential-regime for $\bar{\omega}\leqslant0$ and a damped oscillatory regime for $\bar{\omega}>0$. It's worth noting that the pure oscillatory regime of the ideal fluid Jeans Mechanism is lost. The equation $\bar{\omega}=0$ admits the solutions $K_1$ and $K_2$ which read 
\begin{equation}
K_{\stackrel{\scriptscriptstyle 1}{\scriptscriptstyle 2}}
=\frac{\sqrt{2}\,\ro v_s}{\zetao}
\,\Big(1\mp\sqrt{1-\Big(\frac{K_{J}\zetao}{\ro v_s}\Big)^{2}}\;\;\Big)^{\frac{1}{2}}\;,
\qquad\quad
K_{1},K_{2}>0,\quad K_{1}<K_{2}\;.
\end{equation}
The existence of such solutions gives rise to a constraint on the viscosity coefficient: $\zetao\leqslant\zeta_c=\ro v_s/K_J$. An estimation in the recombination era\footnote{The parameters are set as follows: the usual barotropic relation $p=c^2\ro^\gamma/\tilde{\rho}^{\gamma-1}$ is assumed and the constant $\tilde{\rho}$ can be derived from the expression expression $M_J$ \reff{jeans-mass}. Universe is dominated by matter and we can impose the values: $M_J\sim 10^6 M_\odot$, $\gamma=5/3$, $\rho_{c}=1.879 \,h^2 \cdot 10^{-29} \;g\,cm^{-3}$, $h=0.7$, $z=10^{3}$ and $\ro=\rho_{c}\,\,z^3 = 0.92\cdot10^{-20} \;g\,cm^{-3}$\;. Using these quantities one finds $\tilde{\rho}= 9.034 \cdot 10^{-7}	\;g\,cm^{-3}$, $v_s=8.39\cdot 10^5\;cm\,s^{-1}$ and the threshold value $\zeta_c$.} after decoupling, yields to the value $\zeta_c= 7.38 \cdot 10^{4}	\;g\,cm^{-1}\,s^{-1}$ and confronting this threshold with usual viscosity (\emph{e.g.}, $\zetao^{Hydr.}=8.4\cdot10^{-7}g\,cm^{-1}\,s^{-1}$), we can conclude that the range $\zetao\leqslant\zeta_c$ is the only of physical interest. Finally we obtain: $\bar{\omega}\leqslant0$ for $k\leqslant K_1$, $K_2\leqslant k$ and $\bar{\omega}>0$ for $K_1<k<K_2$.

Let us now analyze the density-perturbation exponential-solutions in correspondence of $\bar{\omega} \leqslant 0$:
\begin{equation}
\delta\rho\sim e^{w\,t}\;,\qquad\quad
w=-\frac{{\zetao}k^2}{2{\ro}}\mp\sqrt{-\bar{\omega}}\;.
\end{equation}
To obtain the structure formation, the amplitude of such stationary waves must grow for increasing time. The exponential collapse for $t\to\infty$ is addressed, choosing the $(+)$ sign solution, only if $w>0$, \emph{i.e.}, $k<K_J$ with $K_J<K_1<K_2$. As a result, we show how the structure formation occurs only if $M>M_J$, as in the standard Jeans Model. 

The viscous effects do not alter the threshold value of the Jeans Mass, but they change the perturbation evolution and the pure oscillatory behavior is lost in presence of dissipative effects. In particular, we get two distinct decreasing regimes: for $K_1<k<K_2$ (\emph{i.e.}, $\wb>0$), we obtain a damped oscillatory evolution of perturbations:
\begin{equation}\label{osc-smorzata}
\delta\rho\sim e^{-\frac{\zeta_0 k^2}{2 {\rho_0}}\,t}\;\cos{(\sqrt{\bar{\omega}}\;t)}\;,
\end{equation}
while, for $K_J<k<K_1$ and $K_2<k$, density perturbations exponentially decrease as $\delta\rho\sim e^{w\,t}$, with $w<0$, in the limit $t\to\infty$.

\subsection{Implication for the top-down mechanism}
As shown above, since the pure oscillatory regime does not occurs, we deal with a decreasing exponential or a damped oscillatory evolution of perturbations. This allows to perform a qualitative analysis of the top-down fragmentation scheme \cite{kolb-turner}, \emph{i.e.}, the comparison between the evolution of two structures: one collapsing agglomerate with $M\gg M_J$ and an internal non-collapsing sub-structure with $M<M_J$. If this picture is addressed, the sub-structure mass must be compared with a decreasing Jeans Mass since the latter is inversely proportional to the collapsing agglomerate background mass. This way, as soon as such a Jeans Mass reaches the sub-structure one, the latter begins to condense implying the fragmentation. In the standard Jeans Model, this mechanism is always allowed since the amplitude for perturbations characterized by $M<M_J$ remains constant in time. On the other hand, the presence of decreasing fluctuations in the viscous model, requires a discussion on the effective damping and an the efficacy of the top-down mechanism. Of course, such an analysis contrasts the hypothesis of a constant background density, but it can be useful to estimate the strength of the dissipative effects. 

We now study two cases for different values of the bulk-viscosity coefficient: $\zetao\ll1$ and $\zetao>1$. In this analysis, a perturbative validity-limit has to be set: we suppose $\delta\rho/\ro\sim0.01$ as the limit of the model and we use the recombination era parameters $^{a}$, in particular the initial time of the collapse is define as the beginning of the matter-dominated Universe, \emph{i.e.}, $t_{\scriptscriptstyle 0}=t_{MD}=1.39\cdot10^{13}\,s$. 

In correspondence of a very small viscosity coefficient (Fig.1),
\begin{figure}[!ht]
\centering
\includegraphics[width=0.9\textwidth]{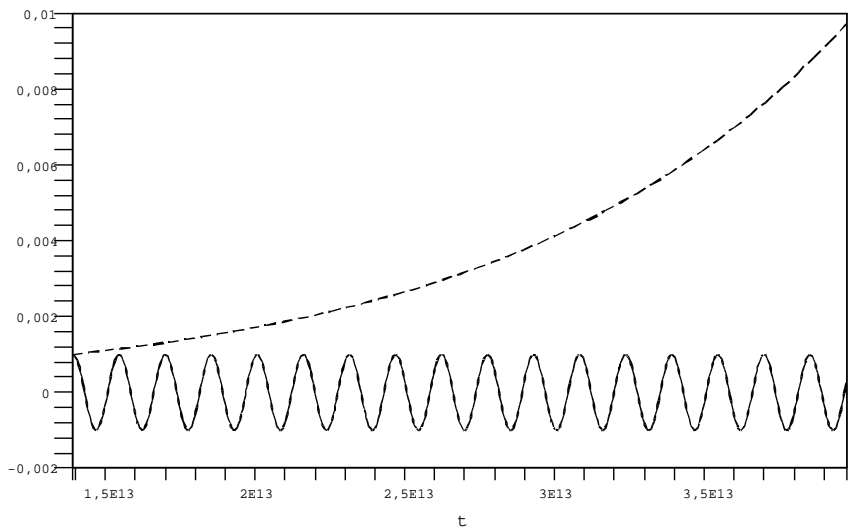}
\caption{\scriptsize{Case $\zetao=10^{-5}\,g\,cm^{-1}\,s^{-1}$.  Galaxy density contrast: $\delta_G$ - $M_G=10^{12}\,M_\odot$ - (dashed line). Sub-structure density contrast $\delta_S$ - $M_S=10\,M_\odot$ - (normal line).}}
\end{figure} 
we consider a decreasing structure of mass $M_S=10\,M_\odot$ within a
collapsing galaxy with mass $M_G=10^{12}\,M_\odot$, the Jeans Mass is $M_J=10^6\,M_\odot$. The sub-structure wave-number $K_S$ is in the region $K_1<K_S<K_2$ and the density perturbations evolve like \reff{osc-smorzata}. Fluctuations have to be imposed small at the initial time $t_{\scriptscriptstyle 0}$, this way, we consider density contrasts ($\delta_G$ for the galaxy and $\delta_S$ for the sub-structure) of $\mathcal{O}(10^{-3})$. In this scheme, the galaxy starts to collapse and the validity limit is reached at $t^*=6.25\cdot10^{13}$. As a result, in Fig.1 we can show how the sub-structure survives in the oscillatory regime during the background collapse until the threshold time value $t^*$. Thus, we can conclude that, if the viscous damping is sufficiently small, the galaxy formation occurs.

Let us now discuss the case $\zetao>1$ (Fig.2) by changing the sub-structure mass, which is now $M_S=M_\odot$.
\begin{figure}[!ht]
\centering
\includegraphics[width=0.9\textwidth]{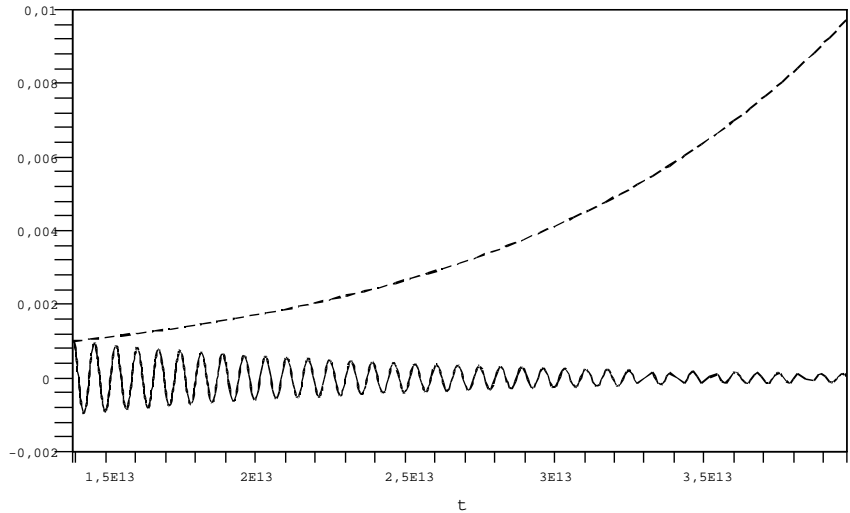}
\caption{\scriptsize{Case $\zetao=14\,g\,cm^{-1}\,s^{-1}$.  Galaxy density contrast: $\delta_G$ - $M_G=10^{12}\,M_\odot$ - (dashed line). Sub-structure density contrast $\delta_S$ - $M_S=M_\odot$ - (normal line).}}
\end{figure}
Here, the viscosity coefficient is greater than one and the damping effects is stronger. In fact, when the galaxy density contrast reaches the threshold value $\delta_G=0.1$, we obtain $\delta_S=10^{-5}$. The top-down mechanism for structure formation results to be unfavored by the presence of strong viscous effects: the damping becomes very strong and the sub-structure vanishes during the agglomerate evolution.

\section{Generalization to the expanding Universe background}
Let us now analyze the behavior of small perturbations, using Newtonian equations, on the expanding-Universe background \cite{weinberg,w71,bonnor57}. The equations which describe the homogeneous- and isotropic-Universe evolution are the well-known Friedmann Eq.'s. Such equations are derived by Einstein Eq.'s using a perfect fluid energy-momentum tensor as the matter source of the gravitational field. It's worth remarking that, we can safely address a Newtonian scheme for astrophysical models, as soon as we treat problems in which the energy density is dominated by non-relativistic particles and in which the linear scales involved are small compared with the characteristic scale of the Universe. Such a matter-dominated Universe is described by the FLRW metric
\begin{equation}\label{metricaFLRW}
ds^2=dt^2 - a^2(t)\,d\ell^2\;,
\end{equation}
with an EoS so that $p\sim0$ ($p\ll\rho$), here $a(t)$ is the scale factor of the Universe.

As in the Jeans Model, we here use the power-law \reff{bulk-power-law} to describe bulk viscosity, which results to be proportional to a positive power of the energy density in the matter-dominated scheme we consider. This way, being the matter density very small, we can consistently neglect viscosity in the unperturbed dynamics. 

The zeroth-order solution corresponds to the evolution of an homogeneous and isotropic Universe filled with the following source: ${T_{\mu}}^{\nu}=\mbox{diag}\,[\,\rho,\,-p,\,-p,\,-p\,]$. The dynamics equations are the energy-momentum conservation-law $T_{\mu;\,\nu}^{\,\nu}=0$ (for $\mu=0$), written in a co-moving frame (here and in the following, the dot ($\dot{\;\;}$) denotes the total derivative wrt time), 
\begin{equation}\label{conservazione-EMT}
\dot{\rho}\,+\, 3\,\frac{\dot{a}}{a}\,\,(\rho+p)= 0\;,
\end{equation}
and the cosmological equation
\begin{equation}\label{eq-cosmologica}
\dot{a}^2 \,+\, {\cal{K}} \,= \,\frac{8 \pi G}{3}\,\rho\, a^2\;,
\end{equation}
where $\mathcal{K}=const.$ is the curvature factor. In this picture, the unperturbed solutions are, setting $\po=0$,
\begin{equation}\label{unperturbed-universe}
\ro= \rb \Big(\;\frac{\ao}{a}\;\Big)^{3}\;,\qquad
\vo = \textbf{r}\,\frac{\dot{a}}{a}\;,\qquad
\nabla\phio=\tfrac{4}{3}\,\textbf{r}\,\pi G\ro\;,
\end{equation}
where $\rb$ and $\ao$ are dimensional constants, $\textbf{r}$ ($r=|\textbf{r}|$) denotes the radial coordinate vector and, of course, $a(t)$ satisfies (\ref{eq-cosmologica}). The solutions $\vo$ and $\phio$ are derived from the Continuity Eq. \reff{continuity-eq} and the Poisson Eq. \reff{poisson-eq} respectively, while the Navier-Stokes Eq. results to be satisfied since the Friedmann Eq.'s hold. 

To obtain the time dependence of the parameter involved in the model, we limit our analysis to early times, since fluctuations arise from the recombination era and, furthermore, the Jeans Mass is so small for recent times that it is of little interest \cite{w71}. This way, the study is restricted to scale factors satisfy the condition $a(t)\ll\ao$, so that ${\dot{a}}^{2},\;8\pi {\rho}a^2/3\gg1$ and we can use the zero-curvature solution without loss of generality. Setting $\mathcal{K}=0$ in the cosmological equation \reff{eq-cosmologica} and using the solution $\ro$ \reff{unperturbed-universe}, one can get the following time dependence
\begin{equation}\label{time-parameters}
a\sim t^{2/3}\;,\qquad
\ro=\frac{1}{6\pi G t^2}\;.
\end{equation}
The study of the gravitational instability is characterized by the evolution of the density contrast and, in particular, of the small fluctuations. In this respect, we underline that $v_s^{2}=\delta p/\delta\rho$ takes account for first-order terms and we have to explicitly write its time dependence during the Universe expansion. For a general specific heat ratio $\gamma$, we can assume that the pressure varies as $\ro^{\gamma}$ and the speed of sound is find to be
\begin{equation}\label{time-vs}
v_s\sim\;t^{1-\gamma}\;.
\end{equation}

Such solutions characterize the background dynamics of the expanding Universe. It's worth noting that, in this generalization, the unperturbed dynamics is now a real solution of the zeroth-order equations and we do not have to apply the ``Jeans swindle'' static-solution assumption.

\subsection{Review of the non-dissipative case}
We want now to study the behavior of the density contrast, without the presence dissipative effects. Since we consider small scales, \emph{i.e.}, $r\ll a$ ($r/a=0)$, as the fluid motion-equations one can assume the system of Newtonian equations \reff{newtonian-motion-eq}. In particular, we neglect the viscous term in equation \reff{navier-stokes-eq} and we perform the usual perturbation theory. 

The resulting first-order motion-equations are spatially homogeneous \cite{weinberg} so one can address plane-wave solutions of the form
\begin{equation}\label{plane-wave-expansion}
\delta\rho(\textbf{r},t) = \ru(t)\,e^{\frac{i\,\textbf{r}\cdot \textbf{q}}{a}}\;,\qquad
\delta\textbf{v}(\textbf{r},t)= \vu(t)\,e^{\frac{i\,\textbf{r}\cdot \textbf{q}}{a}}\;,\qquad
\delta\phi(\textbf{r},t) = \phiu(t)\,e^{\frac{i\,\textbf{r}\cdot \textbf{q}}{a}}\;.
\end{equation}
The factor $1/a(t)$ represents the wave-length reduction dues to the Universe expansion:  $q=|\textbf{q}|$ is the co-moving weave number, being $k=q/a$ the physical one. To complete our analysis, in the limit $r/a=0$, it is convenient to decompose the time depending velocity fluctuations $\vu$ into two part: one transversal and one parallel to the $\textbf{q}$ direction, respectively:
\begin{equation}\label{v-decomposition}
\vu(t)=\vu^\perp\;+\;i\textbf{q}\,\epsilon\;,\qquad
\textbf{q}\cdot\vu^\perp=0\;,\qquad 
\epsilon=-\tfrac{i}{q^2}(\textbf{q}\cdot\vu)\;.
\end{equation}
It is also useful to express $\ru$ in terms of the density contrast: $\ru(t)=\ro\delta$.

A simple algebraic analysis of the first-order dynamics shows that two different types of normal modes arise. The \emph{Rotational Modes} are described by $\vu^\perp$ and simply decay as $\vu^\perp(t)\sim\,1/a$ during the Universe expansion. On the other hand, the \emph{Compressional Modes} are characterized by $\epsilon$ e $\delta$ and require a more interesting analysis. Such modes are described by the equation
\begin{equation}\label{main-eq-weinberg}
\ddot{\delta}\,+\,\frac{2\dot{a}}{a}\,\dot\delta\,+\,
\Big(\frac{v_s^2\,q^2}{a^2}\,-\,4\pi G {\ro}\Big)\,\delta\,=0\;,
\end{equation}
which reduces to the Jeans dispersion-relation \reff{dispersion-jeans} as soon as we set $a=const.$ and consider the physical wave number $k$. Taking into account the zeroth-order time dependence \reff{time-parameters} and \reff{time-vs}, one finds that the density-contrast solution of equation \reff{main-eq-weinberg} involves Bessel functions. Such special functions have different behavior corresponding to small or large proper argument. If the argument is large, \emph{i.e.}, much greater than one, density contrast oscillate, on the other hand, it evolves like
\begin{equation}\label{density-contrast-weinberg}
\delta\sim\,t^{-1/6\pm5/6}\;,
\end{equation}
as soon as the Bessel argument is much less than unity. The condition which separates the two regimes, implying the gravitational collapse in the limit $t\to\infty$ (of course choosing the positive solution for $\delta$), can be write as $v_s^2\,q^2/a^2\,\lesssim\,6\pi G\ro$, which is substantially the same as the Jeans condition derived from \reff{dispersion-jeans}. It worth underling that the standard Jeans condition is perfectly recast if the parameter $\gamma$ of \reff{time-vs} is set to the value $4/3$.

In conclusion, we can infer that the dynamics of the expanding Universe does not modify (substantially) the value of the Jeans Mass which remains the threshold to address the gravitational collapse of structures.

\subsection{Bulk-viscosity effects on the density-contrast dynamics}
We are now aimed at introducing bulk-viscosity effects into the dynamics. As discussed above, such a dissipative effect can be consistently neglected from the zeroth-order analysis, since we are dealing with a matter-dominated Universe. This way, the unperturbed background on which develop the perturbative theory corresponds to the solution \reff{unperturbed-universe} of a Friedmann Universe.

Adding small fluctuations to the Newtonian system \reff{newtonian-motion-eq} and neglecting second-order terms, we get the following set of equations   
\begin{subequations}\label{perturbed-system-1}
\begin{align}
\p_t{\delta\rho}+3\,\frac{\dot{a}}{a}\,\delta\rho
+\frac{\dot{a}}{a}\,(\textbf{r}\cdot\nabla)\delta\rho
+\ro\,\,\nabla\cdot\delta\textbf{v}\,&=0\;,\\ 
\ro\,\p_t{\delta\textbf{v}}+\ro\,\frac{\dot{a}}{a}\,\delta\textbf{v}+
\ro\,\frac{\dot{a}}{a}\,(\textbf{r}\cdot\nabla)\delta\textbf{v}+
v_s^2\;\nabla\,\delta\rho+
\ro\,\nabla\,\delta\phi
-\zetao\;\nabla\,(\nabla\cdot\delta\textbf{v})\,&=0\;,\\
\nabla^2\delta\phi - 4\pi G \delta\rho\,&=0\;,
\end{align}
\end{subequations}
where the relation $\delta p=v_s^2\,\delta\rho$ has been used and we recall that $\zetao=\zeta(\ro)$, see \reff{bulk-expansion}. As in the non-dissipative case, the plane-wave expansion \reff{plane-wave-expansion} for the fluid parameters can be addressed. Using the hypothesis $r/a\sim0$, the system above reduces to:
\begin{subequations}\label{perturbed-system-2}
\begin{align}
\rup+3\,\frac{\dot{a}}{a}\,\ru+\frac{i\,\ro}{a}\,(\textbf{q}\cdot\vu)\,&=0\;,\\
\qquad\qquad\qquad\;
\vup+\frac{\dot{a}}{a}\,\vu
+\frac{i\,v_s^2}{a\,\ro}\;\textbf{q}\,\ru
-4\pi i\,Ga\,\ru\,\frac{\textbf{q}}{q^2}
+\frac{\zetao}{a^ 2\,\ro}\;\textbf{q}\,(\textbf{q}\cdot\vu)\,&=0\;.
\end{align}
\end{subequations}
Let us now follow the standard analysis and use the decomposition \reff{v-decomposition} in order to compare our results wrt the non-dissipative ones. We finally get
\begin{subequations}
\begin{align}
\label{rotational-eq}
\vup^\perp+\frac{\dot{a}}{a}\;\vu^\perp\,&=0\;,\\
\label{compressional-eq1}
\qquad\qquad\qquad\qquad\qquad\qquad\quad\;
\dot{\epsilon}+\Big(\frac{\dot{a}}{a}+\frac{\zetao\,q^2}{\ro\,a^2}\Big)\,\epsilon-
\Big(\frac{4\pi G\ro a}{q^2}- \frac{v_s^2}{a}\Big)\,\delta\,&=0\;,\\
\label{compressional-eq2}
\dot{\delta} -\frac{q^2}{a}\;\epsilon\,&=0\;.
\end{align}
\end{subequations}

The Rotational Modes are not affected by viscosity. In fact, they are governed by equation \reff{rotational-eq} which has the solution 
\begin{equation}
\vu^\perp(t)\sim\,1/a\;,
\end{equation}
as in the standard analysis presented above. On the other hand, the Compressional Modes are influenced by the presence of viscosity. In particular, combining together \reff{compressional-eq1} and \reff{compressional-eq2}, we get an equation which generalizes the compressional equation \reff{main-eq-weinberg}. It reads
\begin{equation}\label{compressional-eq}
\ddot{\delta}\,+\,\Big(2\,\frac{\dot{a}}{a}\,+\,\frac{\zetao q^2}{\ro\,a^2}\Big)
\,\dot\delta\,+\,
\Big(\frac{v_s^2\,q^2}{a^2}\,-\,4\pi G {\ro}\Big)\,\delta\,=0\;.
\end{equation}
This is the fundamental equation which governs the evolutions of the density contrast on an expanding Universe. Let us now write explicitly the time dependence of the parameters involved in the model. The zeroth-order analysis still remains valid in presence of viscosity and we can address expressions \reff{time-parameters} and \reff{time-vs}, as soon as we restrict the study to early times, so that $a(t)\ll\ao$  . Furthermore, using the power-law relation \reff{bulk-power-law} for the bulk-viscosity coefficient, one easily finds
\begin{equation}\label{time-bulk-viscosity}
\zetao=\zetaob\,t^{-2s}\;,\qquad\qquad
\zetaob=z/(6\pi G)^s\;.
\end{equation}

With the help of this expression, we can isolate two constants in the equation \reff{compressional-eq}, which finally rewrites
\begin{equation}\label{density-contrast}
\ddot{\delta}\,+\,\left[\frac{4}{3\,t}\,+\,\frac{\chi}{t^{2(s-1/3)}}\right]\,\dot\delta\,+
\,\left[\frac{\Lambda^2}{t^{2\gamma-2/3}}\,-\,\frac{2}{3\,t^2}\right]\,\delta\,=0\;,
\end{equation}
where the constants $\chi$ and $\Lambda$ are
\begin{equation}
\chi= \frac{t^{2(s-1/3)}\,\zetao q^2}{\ro\,a^2}\;,\qquad\qquad
\Lambda=\frac{t^{\gamma-1/3}\,v_s q}{a}\;.
\end{equation}
This equation can not be analytically solved in general. Following the analysis developed in  \cite{nak-cqg}, let us now discuss the case $s=5/6$. Indeed, this case is the only of physical interest since it deals with the maximum effect that bulk viscosity has without dominating the dynamics, in view of its non-equilibrium perturbative characterization. In fact, in the collapsing limit as $t\to\infty$, if $s>5/6$ the viscous term proportional to $\chi$ results to be of higher order and dominant. On the other hand, it can be neglected in the equation \reff{density-contrast}, if $s<5/6$. Substituting this value in the equation above, one gets the following integrable expression
\begin{equation}\label{density-contrast-5/6}
\ddot{\delta}\,+\,\left[\frac{4}{3}+\chi\right]\,\frac{\dot{\delta}}{t}\,+\,
\left[\frac{\Lambda^2}{t^{2\gamma - 2/3}}\,-\,\frac{2}{3\,t^2}\right]\,\delta\,=0\;.
\end{equation}
The solutions are
\begin{equation}\label{delta1}
\delta(t)=t^{-\frac{1}{6}-\frac{\chi}{2}}\,\,\left[ C_{1}\,J_{n}\Big(\frac{\Lambda t^{-\gb}}{\gb}\Big)\,+\,C_{2}\,Y_{n}\Big(\frac{\Lambda t^{-\gb}}{\gb}\Big)\right]\;,
\end{equation}
where $J_n$ and $Y_n$ denote Bessel functions of first- and second-kind, respectively, and 
\begin{equation}
n=-\sqrt{25+6\chi+9\chi^{2}}\,/\,6\gb\;,\qquad\quad
\gb=\gamma-4/3\;.
\end{equation}

These functions oscillate for $t\ll\Lambda^{1/\gb}$, while for $t\gg\Lambda^{1/\gb}$ the density-contrast solutions \reff{delta1} evolve like
\begin{equation}\label{power-law-density-contrast}
\delta\sim t^{-1/6\,-\,\chi/2\,\,\mp\,\,\gb n}\;.
\end{equation}
A simple analysis of the exponent of such solutions shows how it is always positive, for all values of the viscous parameter $\chi$, as soon as we choose the (-) sign solution. This behavior corresponds to a gravitational collapse, if we consider the asymptotic limit $t\to\infty$. The threshold value which separates the different regimes, implying the growth of the density contrast, is defined by the relation $t>\Lambda^{1/\gb}$ which, using \reff{time-parameters}, corresponds to the Jeans condition \reff{dispersion-jeans}:
\begin{equation}
v_s^2\,q^2/a^2\,\lesssim\,6\pi G\ro\;,
\end{equation}
as in the non-dissipative case. We remark that such solutions will apply only after the recombination, with $4/3<\gamma\leqslant5/3$. In fact, in correspondence of $\gamma=4/3$, the solutions \reff{delta1} show a singular behavior and the equation \reff{density-contrast} requires a different treatment.

As in the standard Jeans Model, the key value of the Jeans Mass is not affected by bulk viscosity, \emph{i.e.}, gravitational collapses for $\delta\to\infty$ are addressed if 
\begin{equation}
k<K_J^*=\sqrt{\frac{6\pi G\ro}{\gb^2v_s^2}}\;.	
\end{equation}
The effect of dissipative processes is to modify the evolution of perturbations. In fact, comparing expression \reff{power-law-density-contrast} wrt the non-dissipative behavior of growing density contrast $\delta\sim t^{2/3}$, see \reff{density-contrast-weinberg}, one can show that the relation $-1/6-\chi/2-\gb n<2/3$ is always verified. We can conclude that the effect of bulk viscosity is to damp the density contrast evolution, suppressing the structure formation as in the Jeans Mechanism.

\section{Concluding remarks}
The effects induced by the presence of bulk viscosity have been analyzed in two different cases in a perturbative scheme: the standard Jeans mechanism and the generalization in treating an expanding Universe background. In both approaches, viscosity has been introduced in the first-order dynamics via a power-low function of the energy density and the effects produced on the unperturbed dynamics have been consistently neglected in view of the phenomenological nature of such kind of viscosity.

The main result, in dealing with the viscous generalization of the models proposed, has been to show how bulk viscosity damps the density contrast evolution maintaining unchanged the threshold value of the Jeans Mass. Such an effect suppresses the sub-structure formation in the top-down fragmentation mechanism. 

In particular, in the analysis of the dissipative Jeans Model, a new decreasing regime for perturbations has been found. The presence of such a behavior allowed the study of the top-down scheme for small and strong viscous effects. In the first case, the density-perturbation amplitude of a sub-structure remains substantially constant during the main structure collapse. On the other hand, if viscous effects are sufficiently strong, the sub-structure vanish in the linear perturbative regime, unfavoring the fragmentation. 

In the second part of this work, the static and uniform background solution for the unperturbed evolution has been generalized by the dynamics of an homogeneous and isotropic Friedmann Universe. In this scheme, a Jeans-like relation has been obtained and a considerable damping of the density contrast growth has been found.

\end{document}